\documentstyle[preprint,aps]{revtex}
\begin{document}
\input epsf.sty
\tightenlines


\bibliographystyle{unsrt}
 
\widetext

\title{Kondo Physics in a Single Electron Transistor}

\author{D. Goldhaber-Gordon,$^{1,2}$ Hadas Shtrikman,$^2$ D.
Mahalu,$^2$ David Abusch-Magder,$^1$ U.~Meirav,$^2$ and M. A. Kastner$^1$} 
\address{$^1$Department of Physics,
Massachusetts Institute of Technology, Cambridge, MA 02139}
\address{$^2$Braun Center for Submicron Research, Department of
Condensed Matter Physics,\\
Weizmann Institute of Science, Rehovot 76100, Israel}
\date{\today}
\maketitle

\date{July 30 1997}
\maketitle
  
\begin{abstract}

How localized electrons interact with delocalized electrons is
a question central to many of the problems at the forefront of solid
state physics. 
The simplest example is called the Kondo phenomenon, and occurs when
an impurity 
atom with an unpaired electron is placed in a metal, and the energy of the
unpaired electron is far below the Fermi
energy.  At low temperatures a spin singlet state is
formed between the unpaired localized electron and delocalized
electrons at the Fermi energy. 

The confined droplet of electrons
interacting with the leads of a single electron transistor (SET) is
closely analogous to an impurity atom interacting with the delocalized
electrons in a metal. \cite{win.meir.lee93} 
We report here measurements on a new
generation of SETs that display all the aspects of the Kondo problem:
the spin singlet forms and causes an enhancement 
of the zero-bias conductance when the number of electrons on the artificial
atom is odd but not when it is even.  The singlet is altered by 
applying a voltage or magnetic field or by increasing the
temperature, all in ways that agree with predictions.\cite{wingreen.kondo}

\end{abstract}
\newpage


When the channel of a transistor is made very small and is
isolated from its leads by tunnel 
barriers it behaves in an unusual way.  A transistor can be thought of as
an electronic switch that is on when it conducts current and off
when it does not.  Whereas a conventional field-effect transistor,
such as one in a 
computer memory, turns on only once when electrons are added to it, the
small transistor turns on and off again every time a single electron is
added to it.\cite{fulton.dolan87,meirav.review}  This increased
functionality may 
eventually make these single electron transistors (SETs)
technologically important. 

The unusual behavior of SETs is a manifestation
of the quantization of charge
and energy caused by the confinement of the droplet of electrons in
the small channel.   
Since similar quantization occurs when electrons are confined in an
atom, the small droplet of electrons is often called an artificial
atom. \cite{kastner.phystoday,ashoori96} Indeed, 
the confined droplet of electrons
interacting with the leads of an SET is closely analogous to an impurity
atom interacting with the delocalized electrons in a metal.
\cite{win.meir.lee93}

Several theoretical papers have predicted that a Kondo singlet could
form in an SET, which would make it possible to study
aspects of the Kondo phenomenon inaccessible
in conventional
systems:  With an SET the number of electrons on the artificial atom can be
changed from odd to even;  the difference in energy between
the localized state and the Fermi energy can be tuned;  the coupling to
the leads can be adjusted; voltage differences can be applied
revealing non-equilibrium Kondo phenomena \cite{wingreen.kondo};
and a single localized state can be studied rather than a statistical
distribution of many impurity states.
However, for
SETs fabricated previously, the binding energy of the spin singlet has
been too
small to observe Kondo phenomena.  Ralph and Buhrman
\cite{ralph.buhr94} have observed the Kondo singlet at a single
accidental  
impurity in a metal point contact, but with only two electrodes and
without control over the structure, they have not been able to observe
all 
the features predicted. We report here measurements on a new
generation of SETs that exhibit all the aspects of the Kondo problem.
To make this report accessible 
to as wide an audience as possible, we discuss here only the
qualitative aspects and leave quantitative comparisons for a later
publication.

We have fabricated SETs using multiple metallic gates (electrodes)
deposited on a 
GaAs/AlGaAs heterostructure containing a 2-dimensional electron gas
(see Figure 1a). First, the electrons are trapped in a plane by
differences in the electronic properties of the heterostructure's
layers.  Second, they are excluded from regions of the plane beneath the 
gates when negative voltages are applied to those gates. This creates
a droplet of electrons separated from the leads by tunnel junctions.
This basic technique has been used previously.
\cite{vanwees.pc,wharam.pc,meirav.prl,johnson,weis.iv}
To make our SETs smaller than earlier ones, we have 
fabricated shallower 2DEG heterostructures \cite{gg.fabrication} as
well as finer metallic gate patterns by electron-beam lithography. The
smaller size of the SETs is critical to our observation of the Kondo
effect. (Dimensions are given in Figure 1a)

Several important energy scales and their relative sizes determine
the behavior of an SET (see Figure 1b). At low temperature, the number
of electrons $N$ on the droplet is a fixed integer (roughly 50 for our
samples). This number may be changed by raising the voltage of a
nearby gate electrode which lowers the energy of
electrons on the droplet relative to the Fermi level in
the leads. The
change in energy necessary to add an electron is
called $U$, and in a simple 
model is the charging energy $e^2/2C$, where $C$ is the capacitance of the
droplet.
Since $U$ is determined by the Coulomb repulsion between pairs of
electrons in the droplet, it scales approximately inversely with the
droplet's radius.

For small droplets, the quantized energy difference between different
spatial electronic states becomes important. We call the typical
energy spacing between spatial states
$\Delta \epsilon$. Another important energy $\Gamma$ is the coupling of
electronic states on the artificial atom to those on the leads,
resulting from tunneling. When $\Gamma$ is made
greater than $\Delta 
\epsilon$, the electrons spread from the artificial atom into the
leads, and quantization of charge and energy is lost, even at $T=0$.

Finally, the energy that determines whether Kondo physics will be
visible is $k T_K$, which is always smaller than
$\Gamma$ ($k$ is Boltzmann's constant and $T_K$ is called the Kondo
temperature).\cite{wingreen.kondo} 
By making smaller SETs we have made 
$\Delta \epsilon$
relatively large, which permits large $\Gamma$ and thus $T_K$ comparable to
accessible temperatures. In semiconductor SETs, $\Gamma$ can be tuned
by changing the voltage on the gates which define the barriers
between artificial atom and leads. We find that with our new
SETs we can vary $\Gamma$ slowly as it approaches $\Delta \epsilon$,
and thus optimize $T_K$. 

In our experiment, we make two types of measurements. In the first, we
apply a voltage of a few $\mu$V between the two leads of the SET, called
the source and drain,
and measure the current that flows through the droplet as a function
of the voltage $V_g$ on one of the SET's gates (the middle electrode
on the left in Fig. 1a). For such small applied
voltage ($< kT/e$), current varies linearly with voltage, and the
zero-bias conductance can be measured.
We use a 10 Hz AC excitation, and perform lock-in
detection of the current. In the second class
of measurements, we add a variable DC offset $V_{ds}$ (up to several mV)
to our AC excitation, and again use lock-in detection of current to
obtain differential conductance $dI/dV_{ds}$ versus $V_{ds}$.

Varying $V_g$ on an SET typically results in adding an
electron to the droplet each time the voltage is increased by a fixed
increment proportional to $U$. Since current can flow through the SET
only when the occupancy of the island is free to fluctuate between $N$
and $N+1$, conductance versus $V_g$ shows a series of
sharp, periodically-spaced peaks.
\cite{meirav.prl,johnson,weis.iv} Figure 2c shows this
behavior when $\Gamma$ is made relatively small. When $\Gamma$ is
large, as in Figs. 2a and 2b, 
we find that these peaks form pairs, with large inter-pair spacing and
small intra-pair spacing. 
The two peaks within a pair have comparable widths and heights,
while between pairs the widths vary significantly.
These observations are direct evidence that two electrons of different
spin are occupying each spatial state. Between paired peaks,
$N$ is odd, while between adjacent pairs it is
even. Since two electrons corresponding to the pair of peaks are added
to the same spatial state, the intra-pair spacing is determined by $U$.
However, when $N$ is even (between pairs) the next electron must be
placed in a different spatial state, so the interpair spacing
is determined by $U + \Delta\epsilon$.

When $\Gamma \le k T$ the peaks in
conductance versus gate voltage become narrower and larger with
decreasing $T$, as illustrated in Fig. 2c,
saturating at a width and height determined by $\Gamma$. This behavior
results from 
the sharpening of the Fermi distribution. \cite{meirav.prl,foxman.prb} Fig. 2
illustrates that this pattern is followed for large $\Gamma$, as well,
on the outside edges of paired peaks at all $T$. However, inside of
pairs the peaks become narrower as $T$ is reduced from 800 mK to 400
mK, but then broaden again at low temperatures, resulting in
increased conductance in the intra-pair valley.
Thus, not only the peak spacing but also the mechanism of conduction itself
is different intra-pair from inter-pair. The enhancement of linear
conductance at low temperature for odd but not even N is a
manifestation of Kondo physics. 
If N is odd there is an unpaired electron with a free spin which can
form a singlet 
with electrons at the Fermi level in the leads.  This coupling results
in an enhanced density of states at the Fermi level in the leads, and 
hence an enhanced conductance.
\cite{nglee.kondo,glazmanraikh.kondo} Raising the temperature
destroys the singlet and attenuates the conductance.

Another aspect of Kondo physics is the sensitivity of the excess
conductance between the pair of peaks to 
the difference in Fermi levels in the two leads. The extra electron in
the droplet couples 
to electrons in both leads, giving an enhanced density of states at 
both Fermi levels. \cite{wingreen.kondo,koenig.kondo,hershfield.kondo}
When the applied voltage is large, separating the Fermi levels in the
two leads, the electrons at the Fermi level in the
higher energy lead can no longer resonantly tunnel into the enhanced
density of states in the lower energy lead, so the extra conductance is
suppressed. This can be seen in Figure 3, a plot of
differential conductance versus $V_{ds}$. The
enhanced conductance is suppressed by a bias of $\sim$ 0.1 mV in either
direction. 

Figure 3 illustrates how this nonlinear conductance measurement also
offers a complementary way of seeing the suppression of the zero-bias
enhancement with temperature. By 600 mK, the Kondo resonance has
almost disappeared completely between the peak pair near $V_g = -70$ mV. 

A magnetic field also alters the Kondo physics. Applying a
magnetic field splits the unpaired localized electron
state into a Zeeman doublet separated by the energy $g \mu_B B$. This
also splits the enhanced density of states at the Fermi level 
into two peaks with energies $g \mu_B B$ above and below the Fermi level.
\cite{wingreen.kondo}
When the Fermi level of one lead is raised or lowered by 
a voltage $g \mu_B B/e$ relative to the other lead, electrons can 
tunnel into the Kondo-enhanced density of states. In
differential conductance at high magnetic field, we thus see 
peaks at $\pm g \mu_B B/e$ (see Figure 3). The splitting of peaks in
differential conductance by {\em twice} $g \mu_B B$ (compared to the spin
splitting of the localized state, $g \mu B$) provides a distinctive
signature of Kondo physics. Since the peaks are broad and overlapping,
the distance between their maxima may underestimate their splitting 
at lower magnetic fields. At 7.5 T, when the peaks no longer overlap,
we find their splitting to be $0.033 \pm .002$ meV/Tesla. In
comparison to measurements on bulk GaAs, this is significantly smaller
than $2 g \mu_B B$ yet larger than $g \mu_B B = 0.025$ meV/Tesla.  
Electron spin resonance measurements have found that spin
splittings in a two-dimensional electron gas are suppressed compared
with values for bulk samples, sometimes by as much as 35\%.
\cite{Dobers88}  Thus, our measurement is consistent with a
splitting of $2 g \mu_B B$.

Figure 4 shows the differential conductance at low
temperature and zero magnetic field as a function of both $V_{ds}$ and
$V_g$. 
This range of $V_g$ spans the better-resolved pair of peaks near $V_g
= -70$ mV in Figure 2 as well as the valleys on either side of it. The
bright diagonal lines result from strong peaks in $dI/dV_{ds}$
(outside the range of $V_{ds}$ shown in Figure 3) marking the
values of $V_{ds}$ and $V_g$ where $N$ can change to $N+1$ or $N-1$. The
slopes of these lines contain information about the relative
capacitances of gates and leads to the droplet of electrons. Together
with the intrapair spacing of peaks in zero-bias conductance versus
$V_g$ at 800 mK (Fig. 2), these relative capacitances give a value for
$U$ of $\approx 0.6$ meV. $\Gamma/\Delta\epsilon$ is of order unity, so
electron wavefunctions in the droplet extend somewhat into the leads,
suppressing 
$U$. When $\Gamma$ is reduced for this same device (as in 
the inset to Figure 2), $U$ increases to $\approx$ 2 meV.  From the difference
between inter- and intra-pair spacing 
in Fig. 2 we find $\Delta \epsilon \sim U$ . To determine
$\Gamma$, we set $V_g$ so that the spatial state corresponding to
the paired peaks is empty. Then the width of the peak in $dI/dV_{ds}$
versus $V_{ds}$ corresponding to tunneling into the empty state
is the bare level width $\Gamma \approx 0.2$ meV. This is 
unmodified by Kondo-enhanced tunneling, which can only occur when the
state is
singly occupied in equilibrium. All three quantities $U,
\Delta\epsilon$ and $\Gamma$ are considerably larger than $kT = 0.0078$ 
meV at base temperature.  

A white vertical line at $V_{ds}=0$ in Figure 4 shows that the
zero-bias conductance is enhanced everywhere between the paired peaks
but not outside the pair. We find, 
in fact, that there is a zero-bias {\em suppression} just outside the
pair, especially to the side toward negative gate voltage where our
Kondo level is unoccupied.\cite{koenig.kondo}

In conclusion, we have observed for the first time all the aspects of
Kondo physics in a single-localized-state system. It is apparent that
the characteristics of our SET are affected dramatically by the strong
coupling of the electron droplet to the leads. In addition to
revealing new physics, this may be important in technologically
relevant devices, which are likely to have even larger values of
$\Gamma$, because $\Gamma$ limits the ultimate speed of such devices. 

We would like to thank G. Bunin for help with fabrication, N. Y.
Morgan for help with measurements, and I.
Aleiner, R. C. Ashoori, M. H. Devoret, D. Esteve, D. C. Glattli, A. S.
Goldhaber, L. Levitov, K. Matveev, N. Zhitenev, and especially N. S.
Wingreen and Y. Meir for valuable discussions. 
This work was supported 
by the U.S. Joint Services Electronics Program under contract from the
Department of the Army Army Research Office. D. G.-G. thanks the
students and staff of the Weizmann Institute's Braun Center for
Submicron Research for their hospitality during his stay there, and
the Hertz Foundation for fellowship support. D. A.-M. thanks the
Lucent Technologies Foundation for Fellowship support. 

\bibliography{main}


\newpage

\begin{figure}
\caption{(a) Scanning electron microscope top view of sample. Three
gate electrodes, the one on the right and the upper and lower ones on
the left, control the tunnel barriers between reservoirs of two-dimensional
electron gas (at top and bottom) and the droplet of electrons. The
middle electrode on the left is used as a gate to change the energy of
the droplet relative to the two-dimensional electron gas. Source and
drain contacts at the top and bottom are not shown. While
the lithographic dimensions of the confined region are 150 nm square,
we estimate that the 
electron droplet has dimensions of 100 nm square due to lateral
depletion. The gate pattern shown was deposited on top of a shallow
heterostructure with the following layer sequence grown on top of a
thick undoped GaAs buffer: 4 nm AlAs, $5$x$10^{12}/$cm$^2$ Si
$\delta$-doping, 1 nm Al$_{.3}$Ga$_{.7}$As, $\delta$-doping, 1 nm 
Al$_{.3}$Ga$_{.7}$As, $\delta$-doping, 5 nm Al$_{.3}$Ga$_{.7}$As, 5 nm
GaAs cap.\protect{\cite{gg.fabrication}} Immediately before
depositing the metal, we etched off the GaAs cap in the areas where
the gates would be deposited, to reduce
leakage between the gates and the electron gas. (b) Schematic energy
diagram of the artificial atom and its leads. The situation shown
corresponds to $V_{ds} < kT/e$, for which the Fermi energies in source
and drain are nearly equal, and to a value of $V_g$ near a conductance
minimum between a pair of peaks corresponding to the same spatial
state. For this case there is and energy cost $\sim U$ to add or
remove an electron. To place an extra electron in the lowest excited
state costs $\sim U + \Delta\epsilon$.
}
\label{fig1}   
\end{figure}
\begin{figure}
\caption{
Temperature dependence of zero-bias conductance through two
different spatial states on the droplet. (a) Paired peaks corresponding to
the two spin states for each spatial state become better resolved with
{\em increasing} temperature from 90 mK to 400 mK. The intrapair
valleys become deeper and the peaks become narrower. (b) From
400 mK to 800 mK the paired peaks 
near $V_g = -70 mV$ broaden. The peaks near $V_g = -25$ mV are still becoming 
better-resolved even at 800 mK since they have larger $\Gamma$ and
hence larger $T_K$. (c) When $\Gamma$ is reduced (as illustrated by
shorter and narrower peaks), $U$ increases 
relative to $\Delta\epsilon$, so peak pairing is no longer evident.
Since the Kondo phenomenon is suppressed, peaks become narrower
as temperature is decreased at all $T$.
}
\label{fig2}
\end{figure}

\begin{figure}
\caption{
Temperature and magnetic field dependence of the zero-bias Kondo
resonance measured in differential conductance. Temperature suppresses
the resonance, while magnetic field causes it to split into a pair of
resonances at finite bias. These scans were made with gate voltage
halfway between the paired peaks near $V_g = -70$ mV in Figure 2.
Since the energy of the spatial state we are studying changes with
magnetic field, the valley between our two peaks occurs at a
different gate voltage for each value of magnetic field.
}
\label{fig3}
\end{figure}

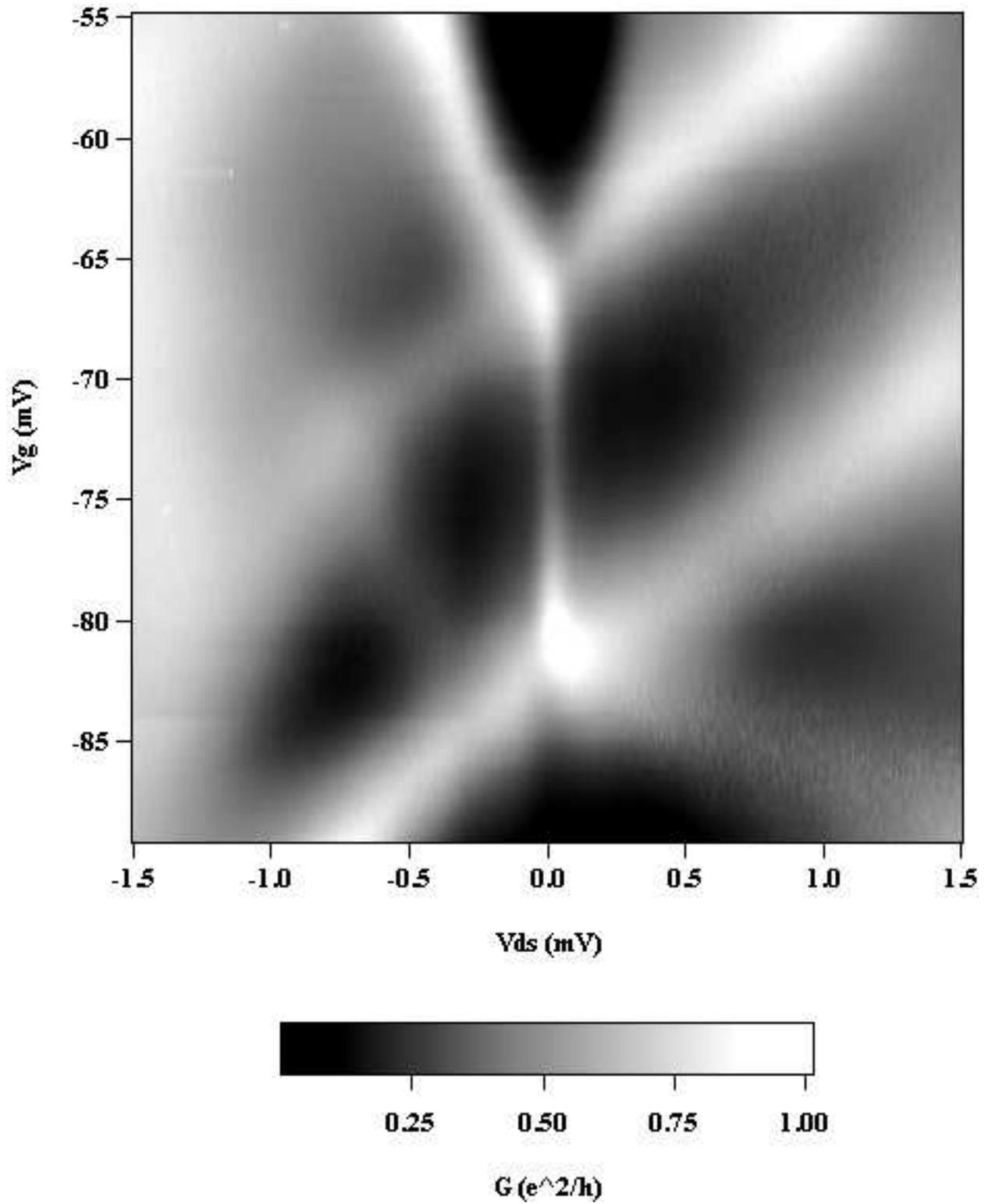
\begin{figure}
\caption{
Differential conductance on a color scale as a function of both $V_g$
and $V_{ds}$. The white vertical line between the two maxima indicates
that there is a zero-bias peak for odd $N$ only.
}
\label{fig4}
\end{figure}

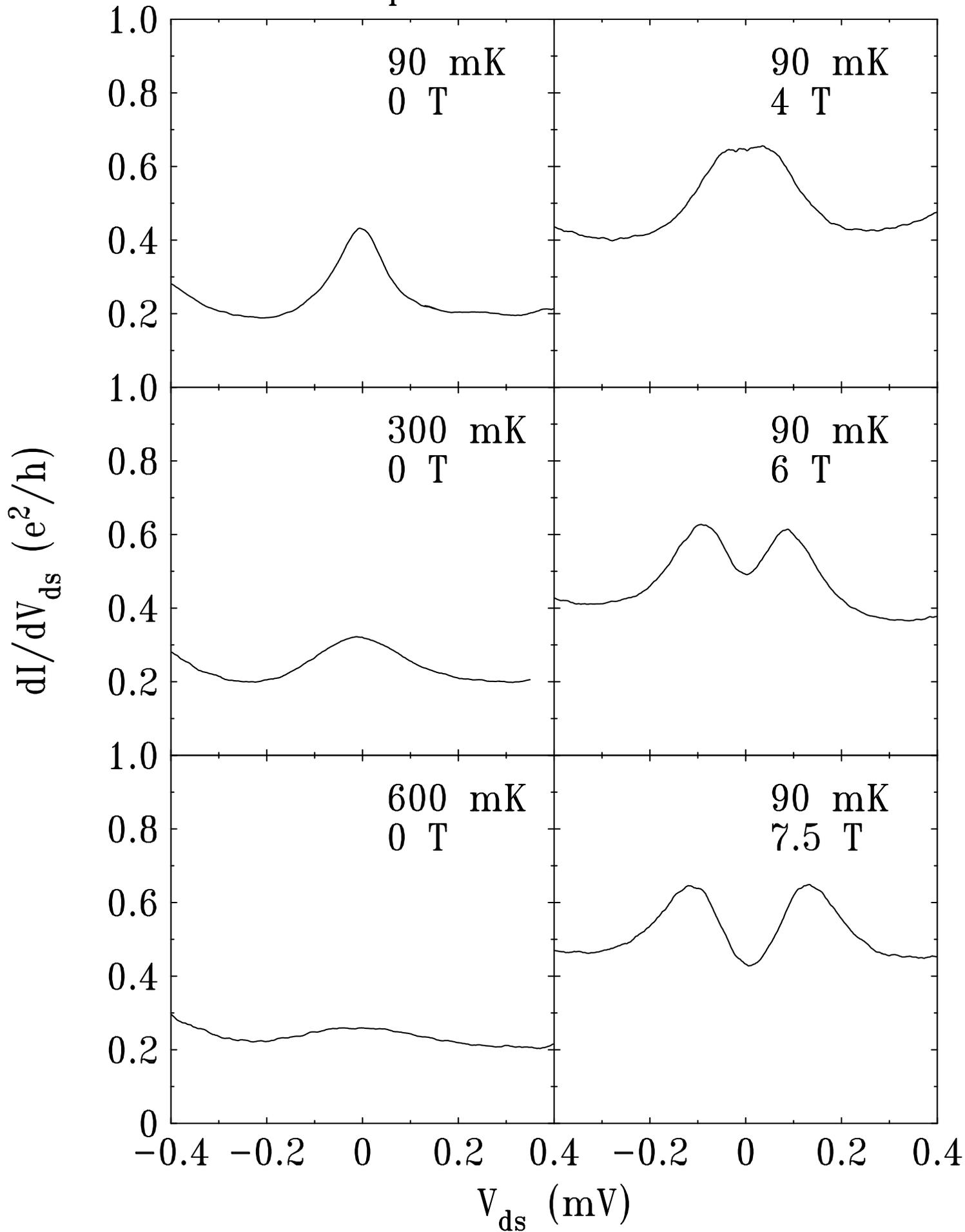
\end{document}